# Triplet fermions in MXenes: The Applications for spintronic-based devices

**Phusit Nualpijit[a,b,*], Bumned Soodchomshom[a]**

[a] Department of Physics, Faculty of science, Kasetsart University, Bangkok,10900, THAILAND
[b] School of Integrated Science (SIS), Kasetsart University, Bangkok,10900, THAILAND

*E-mail: phusit.nual@ku.th

**Abstract**

We investigate the electronic properties of MXenes by three bands tight-binding model of $d_{z^2}$, $d_{xy}$, and $d_{x^2-y^2}$ orbitals. The three corresponding bands touch each other at high symmetry K point in the case of absence of spin-orbit interaction. The proper parameters can be obtained by Slater-Koster parameters related to chemical bonding, $\pi, \sigma$, and $\delta$ bonds. The model calculated for these band structures make an agreement with the same trend as discussed in DFT calculation which the hopping parameters may be identified roughly by fermi velocity. Furthermore, the triplet fermion occurs around K point hosting by flat band, leading to super-Klein tunnelling and anti-super-Klein tunnelling for gapped and gapless pseudospin-1 fermion, respectively. These may apply for nanodevices operated by spin polarization which is more stable than that of the conventional two-dimensional materials.

## 1. INTRODUCTION

An emergence of two-dimensional transition metal carbides, nitride, or carbonitrides (MXenes family) have an impact on the development of nanodevices by means of the dominant electronic properties and optical properties [1, 2]. The unique characteristics such as high surface-to-volume ratio, high ion transport properties, or adjustable surface properties distinguish them from other two dimensional materials [3]. The chemical formula of MXenes can be represented by $MX_2$ which M are the transition metal atoms, such as Sc, Ti, V, Cr, Zr, Mo, Hf, Ta and X = C, N atoms. Actually, the preparation of single layer MXenes can be done by etching A atoms ( A = Al, Si, P, S, Ga, Ge ) of the MAX phases by hydrofluoric acid [4]. The dominance of magnetic moment of transition metal occurs mainly from $d$ – electron. Thus, the theoretical investigation states that the total magnetic moment of transition metal carbides are larger than transition metal nitrides [5] and vanish in both groups for the termination of gas molecules as discussed in Ref.[6]. The adaptable magnetic properties of MXenes suggest applications of energy storage, electromagnetic field shielding, and MXenes-based gas sensing such as force, humidity, temperature sensors [7, 8].

The intrinsic magnetism has been focused because it provides the tuneable magnetic moment by applying the external electric field [9]. The theoretical prediction based on first-principal calculations state that $Ti_2C$ prefers antiferromagnetic phase which has the lowest energy. Furthermore, this phase can be changed to ferrimagnetic semiconductor, half-metal, magnetic metal, non-magnetic metal, and non-magnetic semiconductor by applying the electric field. This goes in the same direction with the report in Ref. [10] that the 1T phase of $Ti_2C$ manifests the antiferromagnetic while 2H phase becomes ferromagnetic which are stable in the room temperature. The plot of spin resolved band structure in Ref. [6] may point out that 1T phase $Ti_2C$ becomes half-metal which has 100% spin polarization around fermi level. These remarkable properties also point out that the spintronic devices can be improved to prevent the unstable spin filtering effect with low power consumption. MXenes is a candidate to improve the technology of multi-value logic (MVL) computing nanodevices [2] which is more stable than that of conventional materials. This steps forward for rapid technologies such as machine learning and artificial intelligence. This evolution of MVL is available to assist the researcher in materials science to acquire the accurate energy band gap and band structure in shorter time [11, 12].

In this paper, the electronic properties of MXenes have been examined by tight-binding model with Slater-Koster parameters. The main orbitals from transition metal atoms lead to three band model corresponding to $d_{z^2}$, $d_{xy}$, and $d_{x^2-y^2}$ orbitals. The tight binding parameters $V_{dd\pi}$, $V_{dd\sigma}$ and $V_{dd\delta}$ can be evaluated by mapping with DFT calculation in Ref.[8] for $Ti_2C$. The pseudospin-1 fermion occurs for a specific condition of the diagonal elements of Hamiltonian. This paves the way to investigate the transmission probability of injected electron through a square potential barrier. Hence, the Klein and anti-Klein tunnelling may be observed by this model within low energy limit around K-point for a **perfect spin polarization** at the fermi level. Furthermore, this model may be adapted in other two-dimensional materials with the similar structure such as MBenes [13].

## 2. MODEL AND FORMALISM

The electronic states in periodic lattice can be described by Bloch function related to translation vectors. This leads to the general formalism of Hamiltonian of the itinerant electrons in materials that can be expressed by

$$H = \sum_{i,\xi} \left( \varepsilon^a_{i\xi} a^\dagger_{i\xi} a_{i\xi} + \varepsilon^b_{i\xi} b^\dagger_{i\xi} b_{i\xi} \right) + \sum_{\langle i,j \rangle, \xi\xi'} \left( t_{ij,\xi\xi'} a^\dagger_{i,\xi} b_{j,\xi'} + h.c. \right)$$
$$+ \sum_{\langle\langle i,j \rangle\rangle, \xi\xi'} \left( t'_{ij,\xi\xi'} a^\dagger_{i,\xi} a_{j\xi'} + t'_{ij,\xi\xi'} b^\dagger_{i,\xi} b_{j,\xi'} \right),$$
(1)

where $a^\dagger_{i\vartheta}(a_{i\vartheta})$ are the creation (annihilation) operator of electron with orbital $\xi$ located at sublattice $a$, atomic site $i$ and on-site energy $\varepsilon^a_{i\xi}$. The hopping parameter $t_{ij,\xi\xi'}$ can be calculated by Slater-Koster parameters related to cosine direction between pair of atoms [14]. The nearest-neighbour vectors can be defined by $\mathbf{a}_1 = a(0,-1)$, $\mathbf{a}_2 = (a/2)(\sqrt{3},1)$ and $\mathbf{a}_3 = (a/2)(-\sqrt{3},1)$. Moreover, the next nearest-neighbour hopping parameter $t'_{ij,\xi\xi'}$ can be acquired by the cosine directions of $\mathbf{c}_i$ which are the combination of $\mathbf{a}_i$.

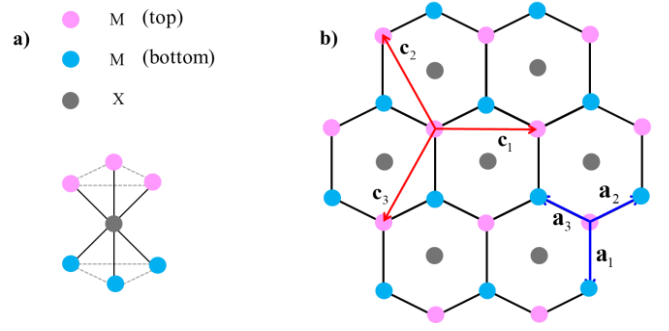

**Fig.1 a)** side view of the T-phase MXenes ($M_2X$) where M represents transition metal atoms such as Sc, Ti, V, Cr and X refers to C, N atoms. **b)** top view of T-phase MXenes with the nearest neighbour vectors $\mathbf{a}_i$ and next-nearest neighbour vectors $\mathbf{c}_i$.

In MXenes, the main effect arises from $d_{z^2}$, $d_{xy}$, and $d_{x^2-y^2}$ orbitals from $M$ atoms which was discussed in DFT calculations [15-17]. The $p$ orbital of $X$ atom can be neglected around fermi level because the band is deeply located below $d$ band of $M$ atom [8]. To acquire the suitable parameters of the tight-binding model related to the correspond orbitals, the basis vectors can be written as

$$\left| d_{z^2} \right\rangle, \quad \left| d_{x^2-y^2} \right\rangle, \quad \left| d_{xy} \right\rangle. \quad (2)$$

Because of the main orbitals that come from $M$ atom, the nearest-neighbour interaction can be ignored. This leads to $3 \times 3$ matrices of Hamiltonian related to $d_{z^2}$, $d_{xy}$, and $d_{x^2-y^2}$ orbitals [17]. The hopping parameters between $M-M$ atoms in the same plane can be evaluated by $t'_{ij,\xi'} \equiv \sum_{\xi\xi',m} f(n_x,n_y,n_z) V_{\xi\xi',m}$ acquired from Slater-Koster method [14] with coefficient $f(n_x,n_y,n_z)$ along unit vector $\mathbf{n}(n_x,n_y,n_z)$ pointing to the next-nearest neighbour of $M$ atoms as in Fig.1b. The standard hopping parameters $V_{\xi\xi',m}$ of orbitals $\xi,\xi'$ and bonding types $m = \pi,\sigma,\delta$ are considered. The spin-orbit interaction may be included in the Hamiltonian because of strong magnetic moment of the transition metal [15, 17, 18] which can be expressed as

$$H_{so} = \frac{\Delta_{so}}{2}\begin{bmatrix} 0 & 0 & 0 \\ 0 & 0 & i \\ 0 & -i & 0 \end{bmatrix}. \quad (3)$$

Thus, the three bands tight-binding Hamiltonian for single layer MXenes can be written by

$$H = \begin{bmatrix} h_0 & h_1 & h_2 \\ h_1{}^* & h_{11} & h_{12} \\ h_2{}^* & h_{12}{}^* & h_{22} \end{bmatrix} \quad (4)$$

where

$$h_0 = \varepsilon_1 + \frac{1}{2}(3V_{dd\delta} + V_{dd\sigma})\left[\cos(\sqrt{3}k_1 a) + 2\cos\left(\frac{\sqrt{3}k_1 a}{2}\right)\cos\left(\frac{3k_2 a}{2}\right)\right]$$

$$h_1 = \frac{\sqrt{3}}{2}(V_{dd\delta} - V_{dd\sigma})\left[\cos(\sqrt{3}k_1 a) - \cos\left(\frac{\sqrt{3}k_1 a}{2}\right)\cos\left(\frac{3k_2 a}{2}\right)\right]$$

$$h_2 = -\frac{3}{4}(V_{dd\delta} - V_{dd\sigma})\left[\cos\left(\frac{\sqrt{3}k_1 a - 3k_2 a}{2}\right) - \cos\left(\frac{\sqrt{3}k_1 a + 3k_2 a}{2}\right)\right]$$

$$h_{11} = \varepsilon_2 + \frac{1}{4}(V_{dd\delta} + 12V_{dd\pi} + 3V_{dd\sigma})\cos\left(\frac{\sqrt{3}k_1 a}{2}\right)\cos\left(\frac{3k_2 a}{2}\right)$$
$$+ \frac{1}{2}(V_{dd\delta} + 3V_{dd\sigma})\cos(\sqrt{3}ka_1)$$

$$h_{12} = -i\Delta_{so} + \frac{\sqrt{3}}{4}(V_{dd\delta} - 4V_{dd\pi} + 3V_{dd\sigma})\sin\left(\frac{\sqrt{3}k_1 a}{2}\right)\sin\left(\frac{3k_2 a}{2}\right)$$

$$h_{22} = \varepsilon_2 + \frac{1}{4}(3V_{dd\delta} + 4V_{dd\pi} + 9V_{dd\sigma})\cos\left(\frac{\sqrt{3}k_1 a}{2}\right)\cos\left(\frac{3k_2 a}{2}\right)$$
$$+ 2V_{dd\pi}\cos(\sqrt{3}k_1 a).$$

The parameter $\varepsilon_1$ is the on-site energy of $d_{z^2}$, and $\varepsilon_2$ for both $d_{x^2-y^2}$ and $d_{xy}$ orbitals [19, 20]. In this case, the flat band around K point occurs with specific condition by substituting the coordinates of K-point into the Hamiltonian. Thus, the condition reads

$$\varepsilon_2 = \varepsilon_1 + \frac{3}{8}V_{dd\sigma} + \frac{3}{2}V_{dd\pi} - \frac{15}{8}V_{dd\delta}. \quad (5)$$

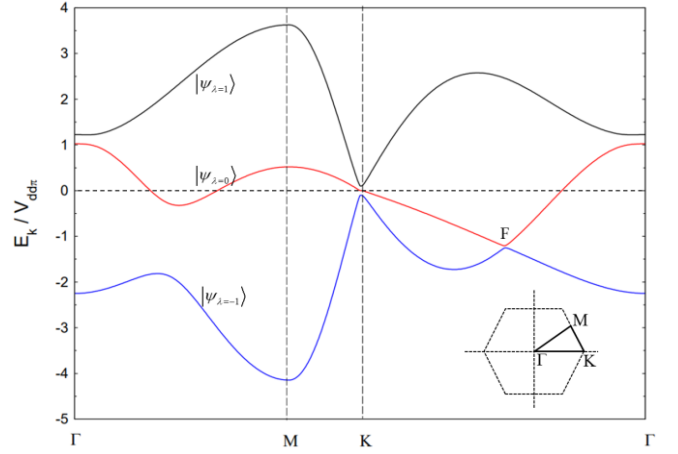

**Fig.2** The band structure of MXenes with $d_{z^2}$, $d_{xy}$, and $d_{x^2-y^2}$ orbitals. These parameters (normalized by $V_{dd\pi}$) are chosen to be $V_{dd\pi}=1$, $V_{dd\sigma}=-1$, $V_{dd\delta}=0$, $\Delta_{so}=0.1$, $\varepsilon_1=-3/4$. The point which the bands cross along $K \to \Gamma$ is called F point [15].

The energy band structure related to $d_{z^2}$, $d_{xy}$, and $d_{x^2-y^2}$ orbitals can be illustrated in Fig.2. This result is in agreement with the same trend of band structure of $Ti_2C$ as illustrated in Ref.[6] which are three bands touching at the K-point when $\Delta_{so}=0$. It is available to transform the Hamiltonian by rotating the axes in Hilbert space which arises from the condition of Eq.5. The unitary operator calculated from K point reads

$$\mathbb{U} = \frac{1}{\sqrt{2}}\begin{bmatrix} 0 & \sqrt{2} & 0 \\ -i & 0 & i \\ 1 & 0 & 1 \end{bmatrix}. \quad (6)$$

This unitary operator satisfies the condition of $\mathbb{U}^\dagger \mathbb{U} = 1$. Therefore, the low energy approximation the Hamiltonian in Eq.4 around K point can be written as

$$H_\eta = \begin{bmatrix} h_0 & h_1 & h_2 \\ h_1{}^* & h_{11} & h_{12} \\ h_2{}^* & h_{12}{}^* & h_{22} \end{bmatrix} \quad (7)$$

which

$$h_0 \approx \varepsilon_0 + \Delta_{SO}$$

$$h_1 \approx i\eta \frac{9\sqrt{3}}{8}(V_{dd\delta} - V_{dd\sigma})\frac{(k_x + ik_y)}{\sqrt{2}}$$

$$h_2 \approx -i\eta \frac{9}{8\sqrt{2}}(V_{dd\delta} - 4V_{dd\pi} + 3V_{dd\sigma})\frac{(k_x - ik_y)}{\sqrt{2}}$$

$$h_{12} \approx i\eta \frac{9\sqrt{3}}{8}(V_{dd\delta} - V_{dd\sigma})\frac{(k_x + ik_y)}{\sqrt{2}}$$

$$h_{11} \approx \varepsilon_0$$

$$h_{22} \approx \varepsilon_0 - \Delta_{SO}.$$

where $\eta = \pm 1$ is the valley index, represented as similar to that of the conventional hexagonal lattice. The eigenvalues of the Hamiltonian at K-point are $E_k = \varepsilon_0, \varepsilon_0 \pm \Delta_{SO}$ where $\varepsilon_0 = -\frac{9}{4}V_{dd\delta} - \frac{3}{4}V_{dd\sigma} + \varepsilon_1$. This leads to the adjustable energy level at K point by $\varepsilon_0$. Thus, the parameter $\varepsilon_1$ in Fig.2 can be evaluated directly by substituting $\varepsilon_0 \to 0$ to acquire the middle band at zero energy level. Efficiently, the fermi velocity can be acquired, the same as in that of conventional hexagonal lattice, which this model gives $v_F \approx \frac{9\sqrt{3}}{8\hbar}V_{dd\sigma}$. The DFT calculation discussed in Ref. [21] has identified the fermi velocity of $Ti_2C$ which is about $0.219 \times 10^6$ m/s. This may specify roughly the tight-binding parameter of $\pi$–bond by $V_{dd\pi} \approx 0.23$ eV for the model parameterized in Fig.2.

## 3. SUPER-KLEIN TUNNELLING OF PSEUDOSPIN-1 FERMION

In this section, we investigate the behavior of electron tunnelling through the square potential barrier. The electron is assumed to be injected with the inciden angle $\theta$ in region I at the position $x = 0$ as shown in Fig.3a. The wave function of electron, in general, may be expressed by

$$|\psi_\lambda\rangle = \frac{1}{2}\begin{pmatrix} -i\alpha\, e^{i\theta} \\ \sqrt{2}\beta \\ i\gamma\, e^{-i\theta} \end{pmatrix}, \qquad (8)$$

with $\alpha = 1 + \lambda\Delta_{SO}/|E|$, $\beta = \lambda\sqrt{1-(\Delta_{SO}/E)^2}$, $\gamma = 1 - \lambda\Delta_{SO}/|E|$, eigenenergies $E = \lambda\sqrt{k^2 + \Delta_{SO}^2}$, and $\lambda = \pm 1$ (neglecting the flat band of E=0 for x<0 due to injected wave requiring a propagating state). Conveniently, this becomes possible by condition $V_{dd\delta} = 0$ and $V_{dd\sigma} = \frac{4}{3}V_{dd\pi}$ for the low energy approximation around nodal point. The wave vectors in $x$ direction of region I, III and II can be written as $k_x = \sqrt{E^2 - \Delta_{SO}^2}\cos\theta$ and $q_x = \sqrt{(E-V_0)^2 - \Delta_{SO}^2}\cos\phi$, respectively. The conservation of momentum in the $y$ direction gives the relationship between incident angle and refracted angle by

$$k_y = q_y \quad \Rightarrow \quad \sqrt{E^2 - \Delta_{SO}^2}\sin\theta = \sqrt{(E-V_0)^2 - \Delta_{SO}^2}\sin\phi. \qquad (9)$$

The solution of eigenvectors can be expressed, in general, by $|\psi\rangle = (\psi_1, \psi_2, \psi_3)^T$ which gives the condition of continuity of $\psi_2$ and $\psi_1 + \psi_3$ at the boundary [22].

The wave function in region I, II, and III as illustrated in Fig.3a-b can be represented by

$$|\psi_{\lambda=1}\rangle_I = \frac{1}{2}\begin{pmatrix} -i\alpha\, e^{i\theta} \\ \sqrt{2}\beta \\ i\gamma\, e^{-i\theta} \end{pmatrix}e^{ik_x x} + r\frac{1}{2}\begin{pmatrix} -i\alpha\, e^{i(\pi-\theta)} \\ \sqrt{2}\beta \\ i\gamma\, e^{-i(\pi-\theta)} \end{pmatrix}e^{-ik_x x}, \qquad (10)$$

$$|\psi_{\lambda=-1}\rangle_{II} = \frac{a}{2}\begin{pmatrix} -i\alpha'\, e^{i\phi} \\ \sqrt{2}\beta' \\ i\gamma'\, e^{-i\phi} \end{pmatrix}e^{iq_x x} + b\begin{pmatrix} -i\alpha'\, e^{i(\pi-\phi)} \\ \sqrt{2}\beta' \\ i\gamma'\, e^{-i(\pi-\phi)} \end{pmatrix}e^{-iq_x x}, \qquad (11)$$

and

$$|\psi_{\lambda=1}\rangle_{III} = \frac{t}{2}\begin{pmatrix} -i\alpha\, e^{i\theta} \\ \sqrt{2}\beta \\ i\gamma\, e^{-i\theta} \end{pmatrix}e^{ik_x x}. \qquad (12)$$

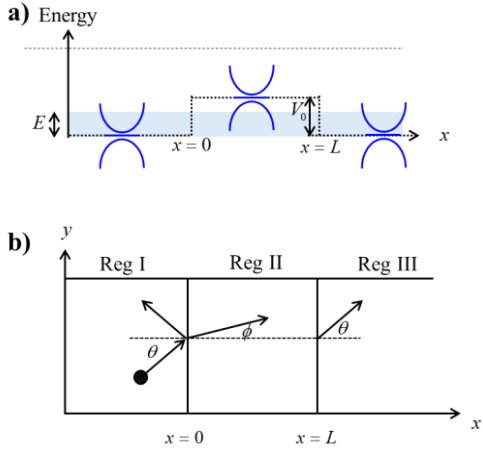

**Fig.3 a)** The Klein tunnelling through a square potential barrier. **b)** The definition of angle $\theta$, $\phi$ in each region.

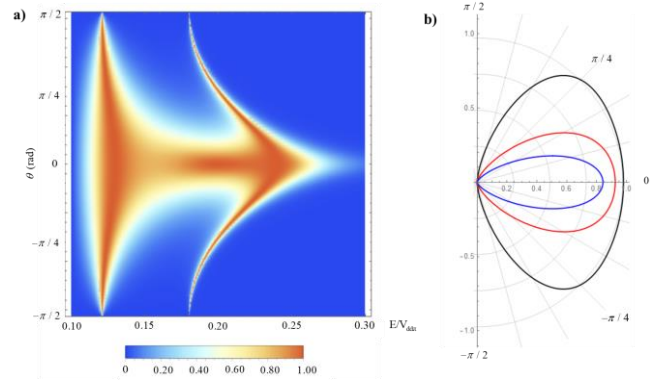

**Fig.4 a)** The transmission probability as function of incident angle $\theta$ and energy E by using $V_0 = 0.4V_{dd\pi}$, $\Delta_{SO} = 0.1V_{dd\pi}$, and $L = 25a$ where $a$ is the lattice constant. **b)** The polar plot of a) with $E/V_{dd\pi} \approx 0.126$ (black line), $E/V_{dd\pi} = 0.14$ (red line), and $E/V_{dd\pi} = 0.16$ (blue line).

The constants $\alpha', \beta', \gamma'$ can be identified as same as in Eq.8 by substituting $E \to E - V_0$. The transmission coefficient $t$ and reflection coefficient $r$ can be evaluated by applying the boundary conditions. This problem can be solved by the system of linear equations for $r, a, b$ and $t$. Finally, the given transmission probability would be defined by the relation $T = |t|^2$.

## 4. RESULTS AND DISCUSSION

Two dimensional layer of transition metal carbides, nitrides and carbonides – known as MXenes – can be investigated by tight-binding model for $d_{z^2}$, $d_{xy}$, and $d_{x^2-y^2}$ orbitals from transition metal atoms which are the main contribution around the fermi level. The $s$ and $p$ orbitals from carbon atoms can be neglected because the bands are located deeply below the fermi level [8, 16, 17]. In this model, the Slater-Koster parameters related to cosine direction are adopted [14]. The $\pi$-bond and $\sigma$-bond are considered, while $\delta$-bond is negligible [23].

The electronic band structure of MXenes can be illustrated by Fig.2 along the line connected with high symmetry points $\Gamma \to M \to K \to \Gamma$ which gives psudospin-1 fermion around the K-point. Futhermore, there is a conical band crossing along $\Gamma \to K$ (F point) when the spin-orbit interaction is absent [15]. In this case, the fermi level has been choosen at the K point (bottom band) which $|\psi_{\lambda=-1}\rangle$ is considered as valence band and $|\psi_{\lambda=0}\rangle, |\psi_{\lambda=1}\rangle$ become two conduction bands. These quantum states operate over the single spin channel around fermi level related to spin filtering effect because the nodal points of each spin polarization lie different energy level [6].

The massive pseudospin-1 fermion can be occurred around the vicinity of the K point which gives rise to the eigenenergies $E_k = 0, \pm\sqrt{v_F^2 k^2 + \Delta_{SO}^2}$ calculated by the effective Hamiltonian around the K point. The investigation of transmission probability of injected electron through a square barrier can be evaluated by using Eq.10-12 which give the relation between incident angle and electron energy as illustrated in Fig.3a-b. Obviously, the transmission probability can be expressed analytically as

$$T = \frac{P}{P\cos^2(q_x D) + Q\sin^2(q_x D)} \quad (13)$$

where

$P = 4\beta\beta'(\alpha-\gamma)^2(\alpha'-\gamma')^2 \cos^2\theta \cos^2\phi$

$Q = \beta'^2(\alpha^2+\gamma^2) + \beta^2(\alpha'^2+\gamma'^2) - 2\alpha\beta'^2\gamma\cos 2\theta$
$\quad - 2\alpha'\beta^2\gamma'\cos 2\phi - 2\beta\beta'(\alpha+\gamma)(\alpha'+\gamma')\sin\theta\sin\phi$

The constants $\alpha, \alpha', \beta, \beta', \gamma, \gamma'$, wave vectors $k_x = \sqrt{E^2 - \Delta_{SO}^2}\cos\theta$, $q_x = \sqrt{(E-V_0)^2 - \Delta_{SO}^2}\cos\phi$, and $\phi = \arcsin\left[\sqrt{E^2 - \Delta_{SO}^2}\sin\theta / \sqrt{(E-V_0)^2 - \Delta_{SO}^2}\right]$ are defined as same as in Eq.8. The relation in Eq.13 expresses the incident angle $\theta$ and energy E dependent of injected electron with fixed potential $V_0$ and spin-orbit coupling strength $\Delta_{SO}$. This can be illustrated by the density plot of transmission probability as shown in Fig.4a. The demonstration of tunnelling electron through the potential barrier is almost completely transmitted when $E/V_{dd\pi} \approx 0.126$ in wide range of incident angle as shown in Fig.4b. On the other hand, the effective model of MXenes offer the distinc results from Dirac model which the transmission probability is absolutely 100% when $E = V_0/2$ and it is found to be incident angle $\theta$ - independent.

For MXenes with massless pseudospin-1 fermion, the spin-orbit interaction in Eq.10-12 has been taken to be zero. The transmission coefficient $t$ vanishes. This suggests the exact solution for anti-super-Klein tunnelling which is omni-direction total reflection for the potential barrier. The outcome become possible in MXenes with gapless because the wave vector $(k_x \pm ik_y)$ has been multiplied by imaginary unit $i$ at the front. This leads to Snell's law calculated by using the boundary $x = 0$,

$$\frac{\sin\theta}{\sin\phi} = -1 \equiv n . \qquad (14)$$

This can be interpreted that the negative refractive index $n$ leads to the antiparallel of pseudospin of incident and transmitted electron at $x = 0$ [24].

The optical response can be investigated by dielectric function and refractive index which is related to the absorption coefficient. The examination of peaks at the critical points can be evaluated by taking the matrix element as a constant [25]. This leads to the join density of states (JDOS) which can be expressed by

$$J_{cv}(E) = -\frac{1}{\pi} \text{Im} \int_{BZ} \frac{1}{E_c - E_v - E + i\delta} \frac{d^2k}{\Omega_{BZ}} . \qquad (15)$$

The expression in Eq.15 has been normalized by the area of Brillouin zone $\Omega_{BZ}$ and included the infinitesimal $\delta$ for applying with Kramers-Konig relation. In this consideration, $|\psi_{\lambda=-1}\rangle$ is the valence band and $|\psi_{\lambda=0}\rangle, |\psi_{\lambda=1}\rangle$ become conduction bands. Thus, the JDOS present a pair of states – occupied state and empty state – separated by $E$. This means that the vertical transition gives $E = E_c - E_v$. The numerical evaluation of JDOS can be illustrated in Fig.5. There are large peaks originated by $\nabla_k(E_c - E_v) = 0$ (critical point). The condition with $\nabla_k E_c = \nabla_k E_v = 0$ occurs at high symmetry points ($\Gamma, M, K$) while $\nabla_k E_c = \nabla_k E_v \neq 0$ occurs in arbitrary point in Brillouin zone [25, 26]. The van Hove singularities are related to the saddle points which the bands are quite flat.

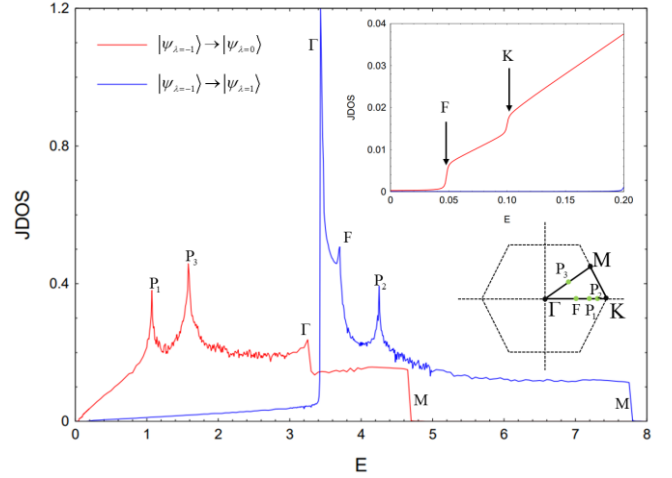

**Fig.5** The join density of states (JDOS) calculated from the band structure in Fig.2. The red line refers to the transition of electron from $|\psi_{\lambda=-1}\rangle \to |\psi_{\lambda=0}\rangle$ and the blue line is for $|\psi_{\lambda=-1}\rangle \to |\psi_{\lambda=1}\rangle$. An inset indicates the low energy limit of JDOS related to the gap of spin-orbit interaction. The van Hove singularities indicated by $P_1, P_2, P_3$ are related to the saddle points of the bands.

The plot can be interpreted that the transition of electron $|\psi_{\lambda=-1}\rangle \to |\psi_{\lambda=0}\rangle$ is dominant when the photon energy is $E/V_{dd\pi} \leq 3$. Furthermore, the conductor ($\Delta_{SO} = 0$) and semiconducting ($\Delta_{SO} \neq 0$) phases may be identified by JDOS as shown in an inset of Fig.5. The threshold energy is related to the energy gap between two corresponding bands. When the gap closes, the JDOS becomes linear relation which is a particular behaviour of hexagonal lattice. This may suggest the suitable range of light frequency for the operation of optical sensing devices [27].

A significant opportunity to accomplish the rapid technologies such as machine learning and artificial intelligence becomes expeditious by muti-value logic computing nanodevices [1, 2]. The half-metallic states of MXenes such as 1T phase $Ti_2C$ become a candidate to prevent the unstable spin-filtering effect because it is conductor in one spin while another spin orientation is semiconductor [6, 28]. This MXenes-based nanodevices may provide an accelerated and convenient way to predict accurately the band gap and tight-binding parameters by machine-learning applied for materials science research [11, 12].

## 5. CONCLUSIONS

We have proposed the model to investigate the electronic properties of transition metal carbides, nitrides, and carbonitrides by tight-binding model. The $d$ – orbitals of transition metal atoms offer the main contribution of band structures near fermi level which are $d_{z^2}$, $d_{xy}$, and $d_{x^2-y^2}$ orbitals. The Slater-Koster parameters related to cosine direction between two atoms suggest the types of bonding, $\pi, \sigma$ and $\delta$ bonds. When the spin-orbit interaction does not exist, three corresponding bands touch at hight symmetry K point by a specific condition in Eq.5. Furthermore, the pseudospin-1 fermion provides the anti-super Klein tunnelling for square potential barrier due to the negative refractive index. This investigation may provide essentially some physical and optical properties for MXenes within half-metallic phases such as $Ti_2C$. This paves the way to develop the rapid technologies operated by spintronics devices such as machine learning and artificial intelligence.

## Acknowledgements

This project is funded by National Research Council of Thailand (NRCT): NRCT5-RSA63002-15. We are also grateful to P. Triwatcharanon for discussing about MXenes in experimental side.